\documentclass[10pt]{article}
\usepackage[]{graphicx}
\usepackage[]{color}
\makeatletter
\def\maxwidth{ %
  \ifdim\Gin@nat@width>\linewidth
    \linewidth
  \else
    \Gin@nat@width
  \fi
}
\makeatother

\definecolor{fgcolor}{rgb}{0.345, 0.345, 0.345}

\usepackage{framed}
\makeatletter
 {\par\unskip\endMakeFramed%
 \at@end@of@kframe}
\makeatother

\definecolor{shadecolor}{rgb}{.97, .97, .97}
\definecolor{messagecolor}{rgb}{0, 0, 0}
\definecolor{warningcolor}{rgb}{1, 0, 1}
\definecolor{errorcolor}{rgb}{1, 0, 0}
\newenvironment{knitrout}{}{} 

\usepackage{alltt}

\usepackage[utf8]{inputenc}
\usepackage{amsmath}
\usepackage{amsfonts}
\usepackage{amssymb}
\usepackage[left=2cm,right=2cm,top=2cm,bottom=2cm]{geometry}
\usepackage{algorithm}
\usepackage{algorithmic}
\usepackage{tikz}
\usepackage{hyperref}
\usepackage{subfig}
\usepackage{graphicx}
\usepackage{epstopdf}
\usepackage{epsfig}
\usepackage{url}
\usepackage{setspace}

\IfFileExists{upquote.sty}{\usepackage{upquote}}{}

\title{Spherical Hamiltonian Monte Carlo for Constrained Target Distributions}
\author{Shiwei Lan\footnote{Department of Statistics, University of California,
Irvine, USA.}, \quad Bo Zhou$^*$, \quad Babak Shahbaba\footnote{Department of
Statistics and Department of Computer Science, University of California, Irvine, USA.}}

\begin{document}


\maketitle
\date

\begin{abstract}
We propose a new Markov Chain Monte Carlo (MCMC) method for constrained target distributions. Our method first maps the $D$-dimensional constrained domain of parameters to the unit ball ${\bf B}_0^D(1)$. Then, it augments the resulting parameter space to the $D$-dimensional sphere, ${\bf S}^D$. The boundary of ${\bf B}_0^D(1)$ corresponds to the equator of ${\bf S}^D$. This change of domains enables us to implicitly handle the original constraints because while the sampler moves freely on the sphere, it proposes states that are within the constraints imposed on the original parameter space. To improve the computational efficiency of our algorithm, we split the Lagrangian dynamics into several parts such that a part of the dynamics can be handled analytically by finding the geodesic flow on the sphere. We apply our method to several examples including truncated Gaussian, Bayesian Lasso, Bayesian bridge regression, and a copula model for identifying synchrony among multiple neurons. Our results show that the proposed method can provide a natural and efficient framework for handling several types of constraints on target distributions.
\end{abstract}


\section{Introduction}
Hamiltonian Monte Carlo (HMC) \cite{duane87,neal10} is a Metropolis algorithm with proposals guided by Hamiltonian dynamics. HMC improves upon random walk Metropolis by proposing states that are distant from the current state, but nevertheless have a high probability of acceptance. These distant proposals are found by
numerically simulating Hamiltonian dynamics, whose state space consists of its \emph{position}, denoted by the vector $\theta$, and its \emph{momentum}, denoted by a vector $p$. Our objective is to sample from the distribution of $\theta$ with the probability density function $p(\theta)$. It is common to assume that the fictitious momentum variable ${p}$ has a multivariate normal distribution with mean zero, ${{p}}\sim N({0, M})$, where $M$ is a symmetric, positive-definite matrix known as the \emph{mass matrix}. In standard HMC, $M$ is usually set to the identity matrix, ${\boldsymbol I}$, for convenience.

Based on $ \theta$ and
$ p$, we define the \emph{potential energy}, $U(\theta)$, and the
\emph{kinetic energy}, $K( p)$. We set $U( \theta)$ to be minus the log probability density of $ \theta$ (plus any constant). For the auxiliary momentum variable $ p$, we set $K(p)$ to be minus the log probability density of $ p$ (plus any constant). The \emph{Hamiltonian} function is then defined as
follows:
\begin{eqnarray}\label{hamiltonian}
H( \theta, p) & = & U(\theta) + K( p)
\end{eqnarray}

The partial derivatives of $H( \theta,  p)$ determine how $ \theta$ and $ p$ change over time, according to \emph{Hamilton's equations},
\begin{eqnarray}\begin{array}{lcrcr }
\displaystyle 
\dot {\theta} & = & \nabla_{ p} H({\theta}, { p}) & = & { M}^{-1}{ p} \\ [12pt]
\displaystyle 
\dot { p} & = & -\nabla_{\theta} H({\theta}, { p})& =  & -\nabla_{\theta} U(\theta) 
\end{array}\label{eq:hmc}\end{eqnarray}
Note that since momentum is mass times velocity, $v = { M}^{-1}{ p}$ is regarded as velocity. Therefore, throughout this paper, we express the kinetic energy in terms of velocity, $v$, as opposed to momentum, $p$ \cite{lan12}. 

Hamiltonian dynamics have three important properties: 1) reversibility (the target distribution remains invariant), 2) conservation of the Hamiltonian (the acceptance probability is one), and 3) volume preservation (the determinant of the Jacobian matrix for the mapping is one). See Neal (2010) \cite{neal10} for more discussion.

In practice, solving Hamilton's equations exactly is difficult, so we need to approximate these equations by discretizing time, using some
small step size $\epsilon$. For this purpose, we could use Euler's method, but it is more common to use the \emph{leapfrog} method, which better approximates Hamiltonian dynamics \cite{neal10}. We can use some number, $L$, of these leapfrog steps, with some
step size, $\epsilon$, to propose a new state in the Metropolis
algorithm. This proposal will be either accepted or rejected based on the Metropolis acceptance probability, which could be less than one.

In recent years, several methods have been proposed to improve the computational efficiency of HMC \cite{beskos11,girolami11,hoffman11,shahbaba13,lan12,byrne13}. In general, these methods do not directly address problems with constrained target distributions. In contrast, in this current paper, we focus on improving HMC-based algorithms when the target distribution is constrained. Neal et~al.~\cite{pneal08}, Sherlock and Roberts \cite{sherlock09}, and Neal and Roberts \cite{pneal12} discuss optimal scaling of random walk Metropolis algorithms when the target distribution is spherically \cite{pneal08} or elliptically constrained \cite{sherlock09}, or when it is confined to a hypercube \cite{pneal12}. When dealing with constrained target distributions, the standard HMC algorithm needs to evaluate each proposal to ensure it is within the boundaries imposed by the constraints. Alternatively, as discussed by Neal \cite{neal10}, one could modify standard HMC such that the sampler bounces back after hitting the boundaries by letting the potential energy go to infinity for parameter values that violate the constraints. This approach, however, is not very efficient computationally. Byrne and \cite{byrne13} discuss a similar approach for distributions defined on a simplex. Brubaker et~al.~\cite{brubaker12} propose a modified version of HMC for handling constraint functions $c(\theta)=0$, and Pakman and Paninski \cite{pakman13} propose an HMC algorithm with an exact analytical solution for truncated Gaussian distributions.  

In this paper, we propose a new Markov Chain Monte Carlo (MCMC) algorithm that provides a natural and efficient framework for sampling from constrained target distributions. Because many types of constraints can be mapped bijectively to the $D$-dimensional unit ball, we first present our method for distributions confined to the unit ball (Section \ref{ball}). The unit ball is a special case of $q$-norm constraints. In Section \ref{otherConst}, we discuss the application of our method for $q$-norm constraints in general. In Section \ref{results}, we evaluate our proposed method using simulated and real data. Finally, we discuss future directions in Section \ref{discussion}.


\section{Sampling from distributions defined on the unit ball} \label{ball}
In many cases, bounded connected constrained regions can be bijectively mapped to the $D$-dimensional unit ball ${\bf B}_0^D(1):=\{\theta\in\mathbb R^D: \Vert \theta\Vert_2
=\sqrt{\sum_{i=1}^D \theta_i^2}\leq 1\}$. Therefore, in this section, we first focus on distributions confined to the unit ball with the constraint $\Vert \theta\Vert_2\leq 1$.

We start by augmenting the original $D$-dimensional parameter $\theta$ with an extra auxiliary variable $\theta_{D+1}$ to form an extended $(D+1)$-dimensional parameter $\tilde \theta = (\theta, \theta_{D+1})$ such that $\Vert \tilde\theta\Vert_2=1$ so $\theta_{D+1} = \pm \sqrt{1-\Vert \theta\Vert_2^2}$. This way, the domain of the target distribution is changed from the unit ball ${\bf B}_0^D(1)$ to the $D$-dimensional sphere, ${\bf S}^D :=\{\tilde \theta\in \mathbb R^{D+1}: \Vert \tilde\theta\Vert_2=1\}$, through the following transformation:
\begin{equation}\label{b2s}
T_{{\bf B}\to {\bf S}}: {\bf B}_0^D(1)\longrightarrow {\bf S}^D, \quad \theta \mapsto \tilde\theta = (\theta, \pm\sqrt{1-\Vert \theta\Vert_2^2})
\end{equation}
Note that although $\theta_{D+1}$ can be either positive or negative, its sign does not affect our Monte Carlo estimates since after applying the above transformation, we can adjust our estimates according to the change of variable theorem as follows:
\begin{equation}\label{domains}
\int_{{\bf B}_0^D(1)} f(\theta) d\theta_{\bf B} = \int_{{\bf S}_+^D} f(\tilde\theta) \left|\frac{d\theta_{\bf B}}{d\tilde\theta_{\bf S}}\right| d\tilde\theta_{\bf S}
\end{equation}
where $\left|\frac{d\theta_{\bf B}}{d\tilde\theta_{\bf S}}\right|=|\theta_{D+1}|$ as shown in Appendix \ref{app:s2b}. Alternatively, we can resample the states according to these weights and use the resulting samples for Monte Carlo estimation and inference. 

\begin{figure}[h]
\begin{center}
\centerline{\includegraphics[width=4in, height=1.9in]{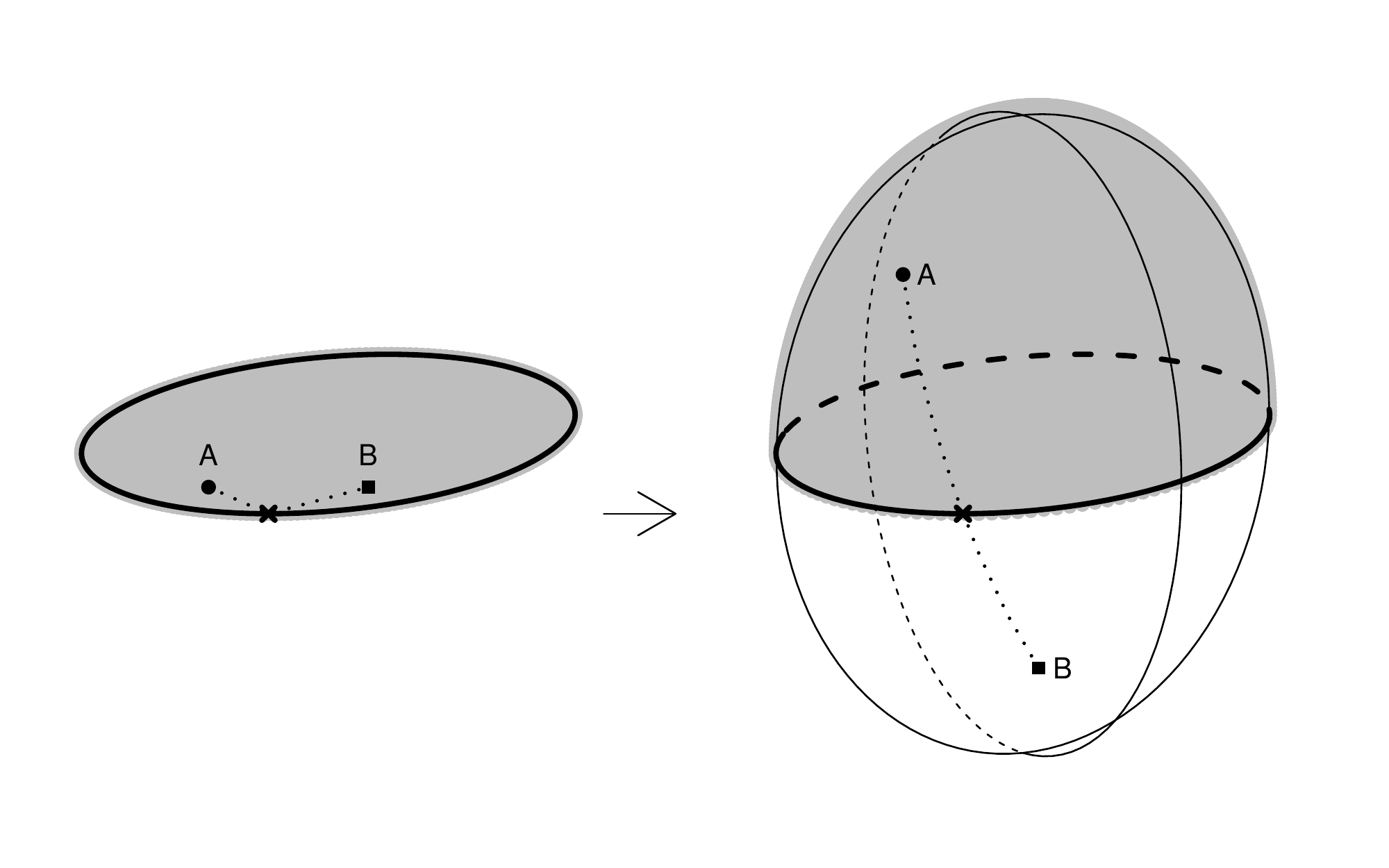}}
\caption{Transforming unit ball ${\bf B}_0^D(1)$ to sphere ${\bf S}^D$.}
\vspace{-15pt}
\label{fig:B2S}
\end{center}
\end{figure}

Using the above transformation, the sampler can move freely on ${\bf S}^D$ implicitly handling the constraints imposed on the original parameters. As illustrated in Figure \ref{fig:B2S}, the boundary of the constraint, i.e., $\Vert\theta\Vert_2=1$, corresponds to the equator on the sphere ${\bf S}^D$. Therefore, as the sampler moves on the sphere, passing across the equator from one hemisphere to the other translates to ``bouncing back'' off the the boundary in the original parameter space. 

In addition to handling the constraint via a simple transformation, our method allows for improving the computational efficiency by using the splitting technique exploited previously by \cite{beskos11, shahbaba13, byrne13}. We consider a family of target distributions, $\{f(\cdot\;;\theta)\}$, defined
on the unit ball ${\bf B}_0^D(1)$ (i.e., the original parameter space) endowed with the Euclidean metric ${\bf I}$. The potential energy is defined as $U(\theta) := -\log f(\cdot\;;\theta)$.
Associated with the auxiliary variable $v$ (i.e., velocity), we define the kinetic energy
$K(v)=\frac{1}{2}v^T{\bf I}v$ for $v\in T_{\theta} {\bf B}_0^D(1)$, which is a $D$-dimensional vector sampled from the tangent space of ${\bf B}_0^D(1)$. Therefore, the Hamiltonian is defined on ${\bf B}_0^D(1)$ as 
\begin{equation}\label{hamiltonianB}
H(\theta,v) = U(\theta) + K(v) = U(\theta) + \frac{1}{2}v^T{\bf I}v 
\end{equation}

Next,  we derive the corresponding Hamiltonian function on ${\bf S}^D$. The potential energy
$U(\tilde\theta)=U(\theta)$ remains the same since the distribution is fully defined in terms of the original parameter $\theta$, i.e., the first $D$ elements of $\tilde \theta$. However, the kinetic energy, $K(\tilde v) := \frac{1}{2} \tilde v^T \tilde v$, changes since the velocity $\tilde v = (v, v_{D+1})$ is now sampled from the tangent space of sphere, $T_{\tilde\theta}{\bf S}^D:=\{\tilde v\in \mathbb
R^{D+1}|\tilde\theta^T \tilde v=0\}$, with $v_{D+1}=-\theta^T
v/\theta_{D+1}$. Therefore, on the sphere ${\bf S}^D$, the Hamiltonian $H^*(\tilde\theta, \tilde v)$ is
defined as follows:
\begin{equation}\label{hamiltonianS}
H^*(\tilde\theta, \tilde v) = U(\tilde\theta) + K(\tilde v)
\end{equation}

Viewing $\{\theta, {\bf B}_0^D(1)\}$ as a coordinate chart of ${\bf S}^D$,
this is equivalent to replacing the Euclidean metric ${\bf I}$ with the \emph{canonical
spherical metric} ${\bf G}_{\bf S}=I_{D}+\theta\theta^T/(1-\Vert\theta\Vert_{2}^2)$. Therefore, we can write the Hamiltonian function as
\begin{equation}\label{hamiltonianSB}
H^*(\tilde\theta, \tilde v) = U(\tilde\theta) + \frac{1}{2} \tilde v^T \tilde v = U(\theta) +
\frac{1}{2}v^T {\bf G}_{\bf S}v
\end{equation}
More details are provided in Appendix \ref{geomS}.

For the above dynamics, we can sample the velocity $v\sim \mathcal N(0,{\bf G}_{\bf S}^{-1})$ and set $\tilde v=\begin{bmatrix}I \\-\theta^T /\theta_{D+1}\end{bmatrix} v$. Alternatively, we can sample $\tilde v$ directly from the standard $D+1$-dimensional Gaussian as follows: 
\begin{equation}\label{VonSph}
\tilde v \sim \mathcal N\left(0, \begin{bmatrix}I \\-\theta^T /\theta_{D+1}\end{bmatrix} {\bf G}_{\bf S}^{-1} \begin{bmatrix}I & -\theta/\theta_{D+1}\end{bmatrix} \right) = \mathcal N(0, I_{D+1} - \tilde\theta \tilde\theta^T ) =  (I_{D+1} - \tilde\theta \tilde\theta^T) \mathcal N(0, I_{D+1})
\end{equation}

The Hamiltonian function \eqref{hamiltonianSB} can be used to define an HMC on the Riemannian manifold $({\bf B}_0^D(1), {\bf G}_{\bf S})$. Equivalently, we can rewrite the above dynamics as the following Lagrangian dynamics \cite{lan12}:
\begin{eqnarray}\begin{array}{lcl}
\dot \theta & = & v\\
\dot v & = & -v^T \Gamma v - {\bf G}_{\bf S}^{-1} \nabla U(\theta)
\end{array}\label{sphLD}\end{eqnarray}
where $\Gamma$ are the Christoffel symbols of second kind derived from ${\bf G}_{\bf S}$.
The Hamiltonian \eqref{hamiltonianSB} is preserved under Lagrangian dynamics \eqref{sphLD}. (See \cite{lan12} for more discussion.)

Following \cite{byrne13}, we could split the Hamiltonian \eqref{hamiltonianSB} as follows:
\begin{equation}
H^*(\tilde\theta, \tilde v) = U(\theta)/2 + \frac{1}{2}v^T {\bf G}_{\bf S}v + U(\theta)/2
\end{equation}
Here, however, we propose an alternative approach based on splitting the Lagrangian dynamics \eqref{sphLD} into two parts, corresponding to $U(\theta)/2$ and $\frac{1}{2}v^T {\bf G}_{\bf S}v$ respectively, as follows:
\begin{eqnarray}
\left\{\begin{array}{lcl}
\dot \theta & = & 0\\
\dot v & = & -\frac{1}{2} {\bf G}_{\bf S}^{-1} \nabla U(\theta)
\end{array}\right.\label{sphLD:U}
\qquad
\left\{\begin{array}{lcl}
\dot \theta & = & v\\
\dot v & = & -v^T \Gamma v
\end{array}\right.\label{sphLD:K}
\end{eqnarray}
(See Appendix \ref{splitHS} for more details.) Note that the first dynamics (on the left) only involves updating velocity $\tilde v$ in the tangent space $T_{\tilde\theta}{\bf S}^D$ and has the following solution
(see Appendix \ref{splitHS} for more details):
\begin{eqnarray}\begin{array}{lcl}
\tilde\theta(t) & = & \tilde\theta(0)\\
\tilde v(t) & = & \tilde v(0) -\frac{t}{2}\left(\begin{bmatrix} I_D\\ 0\end{bmatrix} -\tilde\theta(0) \theta(0)^T\right) \nabla U(\theta(0))
\end{array}\label{evolv}\end{eqnarray}
where $t$ denotes time. 

The second dynamics (on the right) only involves the kinetic energy; hence, it is
equivalent to the geodesic flow on the sphere ${\bf S}^D$ with a \emph{great circle} (orthodrome or Riemannian circle) as its analytical solution (see Appendix \ref{GEODS} for
more details),
\begin{eqnarray}\begin{array}{lcl}
\tilde\theta(t) & = & \tilde\theta(0) \cos(\Vert \tilde v(0)\Vert_{2} t) + \frac{\tilde v(0)}{\Vert \tilde v(0)\Vert_{2}} \sin(\Vert \tilde v(0)\Vert_{2} t)\\
\tilde v(t) & = & -\tilde\theta(0) \Vert \tilde v(0)\Vert_{2} \sin(\Vert \tilde v(0)\Vert_{2} t) + \tilde v(0) \cos(\Vert \tilde v(0)\Vert_{2} t)
\end{array}\label{gcir}\end{eqnarray}

Note that \eqref{evolv} and \eqref{gcir} are both symplectic. Due to the explicit formula for the geodesic flow on sphere, the second dynamics in \eqref{sphLD:K} is simulated exactly. Therefore, updating $\tilde\theta$ does not involve discretization error so we can use large step sizes. This could lead to improved computational efficiency. Since this step is in fact a rotation on sphere, we set the trajectory length to be $2\pi/D$ and randomize the number of leapfrog steps to avoid periodicity. Algorithm \ref{Alg:SphHMC} shows the steps for implementing this approach, henceforth called \emph{Spherical HMC}. 

\begin{algorithm}[t]
\caption{Spherical HMC}
\label{Alg:SphHMC}
\begin{algorithmic}
\STATE Initialize $\tilde\theta^{(1)}$ at current $\tilde\theta$ after transformation
\STATE Sample a new momentum value $\tilde v^{(1)}\sim \mathcal N(0,I_{D+1})$
\STATE Set $\tilde v^{(1)} \leftarrow \tilde v^{(1)} - \tilde\theta^{(1)} (\tilde\theta^{(1)})^T \tilde v^{(1)}$
\STATE Calculate $H(\tilde\theta^{(1)},\tilde v^{(1)})=U(\theta^{(1)}) + K(\tilde v^{(1)})$ for the current state
\FOR{$\ell=1$ to $L$}
\STATE $\tilde v^{(\ell+1/2)} = \tilde v^{(\ell)}-\frac{\epsilon}{2} \left(\begin{bmatrix} I_D\\ 0\end{bmatrix} -\tilde\theta^{(\ell)} (\theta^{(\ell)})^T\right) \nabla U(\theta^{(\ell)})$
\STATE $\tilde \theta^{(\ell+1)} = \tilde \theta^{(\ell)} \cos(\Vert \tilde v^{(\ell+1/2)}\Vert_{2} \epsilon) + \frac{\tilde v^{(\ell+1/2)}}{\Vert \tilde v^{(\ell+1/2)}\Vert_{2}} \sin(\Vert \tilde v^{(\ell+1/2)}\Vert_{2} \epsilon)$
\STATE $\tilde v^{(\ell+1/2)} \leftarrow -\tilde\theta^{(\ell)}\Vert \tilde v^{(\ell+1/2)}\Vert_{2}\sin(\Vert \tilde v^{(\ell+1/2)}\Vert_{2} \epsilon) + \tilde v^{(\ell+1/2)}\cos(\Vert \tilde v^{(\ell+1/2)}\Vert_{2} \epsilon)$
\STATE $\tilde v^{(\ell+1)} = \tilde v^{(\ell+1/2)}-\frac{\epsilon}{2} \left(\begin{bmatrix} I_D\\ 0\end{bmatrix} -\tilde\theta^{(\ell+1)} (\theta^{(\ell+1)})^T\right) \nabla U(\theta^{(\ell+1)})$
\ENDFOR
\STATE Calculate $H(\tilde\theta^{(L+1)},\tilde v^{(L+1)})=U(\theta^{(L+1)}) + K(\tilde v^{(L+1)})$ for the proposed state
\STATE Calculate the acceptance probability $\alpha = \exp\{-H(\tilde\theta^{(L+1)},\tilde v^{(L+1)})+H(\tilde\theta^{(1)},\tilde v^{(1)})\}$
\STATE Accept or reject the proposal $(\tilde \theta^{(L+1)},\tilde v^{(L+1)})$ according to $\alpha$
\STATE Calculate the corresponding weight $|\theta_{D+1}^{(n)}|$
\end{algorithmic}
\end{algorithm}


\section{Norm constraints} \label{otherConst}

The unit ball region discussed in the previous section is in fact a a special case of $q$-norm constraints. In this section we discuss $q$-norm constraint in general and show how they can be transformed to the unit ball so that the Spherical HMC method can still be used. In general, these constraints are expressed in terms of $q$-norm of parameters, 
\begin{equation}\label{qnorm}
\Vert \beta\Vert_q = \left\{
\begin{array}{ll}
(\sum_{i=1}^D |\beta_i|^q)^{1/q}, & q\in (0,+\infty)\\
\max_{1\leq i\leq D} |\beta_i|, & q=+\infty
\end{array}\right.
\end{equation}
For example, when $\beta$ are regression parameters, $q=1$ corresponds to Lasso method, and $q=2$ corresponds to ridge regression. In what follows, we show how this type of constraints can be transformed to ${\bf S}^D$.

\subsection{Norm constraints with $q = +\infty$}\label{sec:qInfinity}
When $q = + \infty$, the distribution is confined to a hypercube. Note that hypercubes, and in general hyper-rectangles, can be transformed to the unit hypercube, ${\bf C}^D:=[-1,1]^D=\{\beta \in\mathbb R^D: \Vert \beta \Vert_{\infty} \leq 1\}$, by proper shifting and scaling of the original parameters. Neal \cite{neal10} discusses this kind of constraints, which could be handled by adding a term to the energy function such that the energy goes to infinity for values that violate the constraints. This creates ''energy walls'' at boundaries. As a result, the sampler bounces off the energy wall whenever it reaches the boundaries. Throughout this paper, we refer to this approach as \emph{Wall HMC}. 

The unit hypercube can be transformed to its inscribed unit ball throughout the following map:
\begin{equation}\label{cuball}
T_{{\bf C}\to {\bf B}}: [-1,1]^D \rightarrow {\bf B}_0^D(1), \quad \beta \mapsto \theta = \beta \frac{\Vert \beta\Vert_{\infty}}{\Vert \beta\Vert_2} 
\end{equation}
Further, as discussed in the previous section, the resulting unit ball can be mapped to sphere ${\bf S}^D$ through $T_{{\bf B}\to {\bf S}}$ for which the Spherical HMC can be used. See Appendix \ref{transJ} for more details.

%

\subsection{Norm constraints with $q\in (0,+\infty)$}\label{sec:qnorm}
A domain constrained by $q$-norm 
${\bf Q}^D:= \{x\in \mathbb R^D: \Vert \beta\Vert_q \leq 1\}$
for $q\in (0,+\infty)$ can be transformed to the unit ball ${\bf B}_0^D(1)$ via
the folllowing map:
\begin{equation}\label{qball}
T_{{\bf Q}\to {\bf B}}: {\bf Q}^D \rightarrow {\bf B}_0^D(1), \quad \beta_i \mapsto \theta_i = \mathrm{sgn}(\beta_i)|\beta_i|^{q/2}
\end{equation}
As before, the unit ball can be transformed to sphere for which we can use the Spherical HMC method. More details are provided in
Appendix \ref{transJ}.

%

\section{Experimental results} \label{results}
In this section, we evaluate our proposed methods,
Spherical HMC, by comparing its efficiency to that of Random Walk Metropolis (RWM) and Wall HMC using simulated and real data. To this end, we define efficiency in terms of time-normalized effective sample size (ESS). Given $B$ MCMC samples for each parameter, we calculate the corresponding ESS = $B[1 + 2\Sigma_{k=1}^{K}\gamma(k)]^{-1}$, where $\Sigma_{k=1}^{K}\gamma(k)$ is the sum of $K$ monotone sample autocorrelations \cite{geyer92}. We provide {minimum, median, and maximum} values of ESS
over all parameters. However, we use the minimum ESS normalized by the CPU time, s (in seconds), as the overall measure of efficiency: $\min(\textrm{ESS})/\textrm{s}$. All computer codes are available online at \url{http://www.ics.uci.edu/~slan/lanzi/CODES.html}.

\subsection{Truncated Multivariate Gaussian}
For illustration purposes, we first start with a truncated bivariate Gaussian distribution,
\begin{eqnarray*}
\binom{\beta_1}{\beta_2} \sim \mathcal N\left(\bf{0}, \begin{bmatrix} 1& .5\\ .5 & 1\end{bmatrix} \right), \\
\\
0\leq \beta_1\leq 5, \qquad 0\leq \beta_2\leq 1
\end{eqnarray*}
The lower and upper limits are $l=(0,0)$ and $u=(5,1)$ respectively. The original rectangle
domain can be mapped to the 2-dimensional unit sphere through the following transformation:
\begin{equation}
T: [0,5]\times [0,1] \rightarrow S^2, \quad \beta \mapsto \beta' = (2\beta-(u+l))/(u-l) \mapsto \theta= \beta' \frac{\Vert \beta'\Vert_{\infty}}{\Vert \beta'\Vert_2} \mapsto \tilde\theta = \big (\theta,\sqrt{1-\Vert \theta\Vert_2^2} \, \big)
\end{equation}

The left panel of Figure \ref{fig:TMG-sph} shows the heatmap based on the exact density funtion, and the right panel shows the corresponding heatmap based on MCMC samples from Spherical HMC. Table \ref{TMG-moments} compares the true mean and covariance of the above truncated bivariate Gaussian distribution with the point estimates obtained from RWM, Wall HMC, and Spherical HMC using 100000 MCMC iterations. Overall, all methods provide reasonably well estimates. 

\begin{knitrout}
\definecolor{shadecolor}{rgb}{0.969, 0.969, 0.969}\color{fgcolor}\begin{figure}[htbp]

{\centering \includegraphics[width=.8\textwidth,height=.4\textwidth]{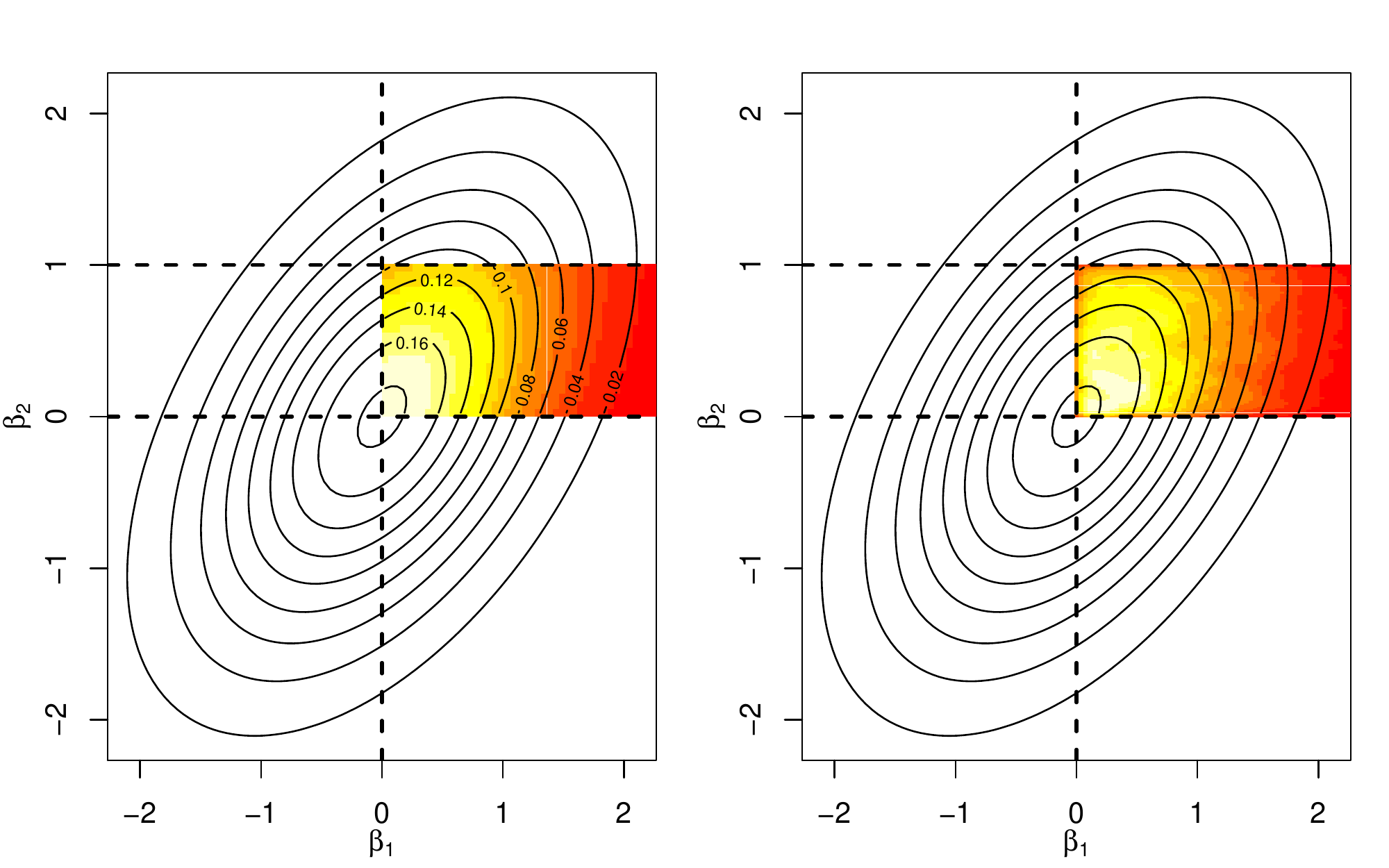} 

}

\caption[Truncated Multivariate Gaussian]{Density plots of a truncated bivariate Gaussian using exact density function (left) and MCMC samples from Spherical HMC (Right). \label{fig:TMG-sph}}
\end{figure}

\end{knitrout}

\begin{table}[ht]
\centering
\begin{tabular}{l|c|c}
  \hline
Method & Mean & Covariance \\
  \hline
Truth & $\begin{pmatrix}0.791\\0.489\end{pmatrix}$ & $\begin{pmatrix}0.327&0.017\\0.017&0.080\end{pmatrix}$ \\ [2.1ex]
   \hline
RWM & $\begin{pmatrix}0.776\\0.489\end{pmatrix}$ & $\begin{pmatrix}0.322&0.015\\0.015&0.080\end{pmatrix}$ \\ [2.1ex]
   \hline
Wall HMC & $\begin{pmatrix}0.793\\0.489\end{pmatrix}$ & $\begin{pmatrix}0.328&0.016\\0.016&0.080\end{pmatrix}$ \\ [2.1ex]
   \hline
Spherical
HMC & $\begin{pmatrix}0.792\\0.489\end{pmatrix}$ & $\begin{pmatrix}0.326&0.017\\0.017&0.079\end{pmatrix}$ \\ [2.1ex]
   \hline
\end{tabular}
\caption{Comparing the point estimates of mean and covariance matrix of a bivariate truncated Gaussian distribution using RWM, Wall HMC, and Spherical HMC.} 
\label{TMG-moments}
\end{table}

To evaluate the efficiency of the above three methods (RWM, Wall HMC, and Spherical HMC), we repeat the this experiment for higher dimensions, $D=10$, and $D=100$. As before, we set the mean to zero and set the $(i,j)$-th element of the covariance matrix to 
$\Sigma_{ij}=1/(1+|i-j|)$. Further, we impose the following constraints on the parameters, 
\begin{equation}
0\leq \beta_i\leq u_{i}
\end{equation}
where $u_{i}$ (i.e., the upper bound) is set to 5 when $i=1$; otherwise, it is set to $0.5$.

For each method, we obtain 10000 MCMC samples after discarding the initial 1000 samples. We set the tuning parameters of algorithms such that their overall acceptance rates are within a reasonable range. For RWM, about $95\%$ of times proposed states are rejected due to violating the constraints. 
Wall HMC improves over RWM, but its efficiency is negatively affected by the computational overhead of monitoring hitting the energy wall, which requires evaluating the boundary conditions and the distance between the proposed state from the boundary. On average, Wall HMC bounces off the wall around 7.68 and 31.10 times per iteration for $D=10$ and $D=100$ respectively. In contrast, by augmenting the parameter space, Spherical HMC handles the constraints in an efficient way. As shown in Table \ref{TMG-eff}, its overall efficiency (measured in terms of time-normalized minimum effective sample size) is substantially higher than that of RWM and Wall HMC. 

\begin{table}[ht]
\centering
\begin{tabular}{l|l|cccc}
  \hline
Dim & Method & AP & s & ESS & Min(ESS)/s \\ 
  \hline
& RWM & 0.64 & 1.59E-04 & (15,75,91) & 8.80 \\ 
  D=10 & Wall HMC & 0.93 & 5.81E-04 & (2725,7738,8376) & 426.79 \\ 
   & Spherical HMC & 0.81 & 9.73E-04 & (6455,8220,8578) & 602.78 \\ 
   \hline
 & RWM & 0.72 & 1.28E-03 & (1,4,18) & 0.06 \\ 
  D=100 & Wall HMC & 0.94 & 1.39E-02 & (2175,6900,7691) & 14.23 \\ 
   & Spherical HMC & 0.88 & 1.51E-02 & (6680,8855,10000) & 40.12 \\ 
   \hline
\end{tabular}
\caption{Sampling Efficiency in of RWM, Wall HMC, and Spherical HMC for generating samplers from truncated Gaussian distributions.} 
\label{TMG-eff}
\end{table}

\subsection{Bayesian Lasso}
In regression analysis, overly complex models tend to overfit the data. Regularized regression models control complexity by imposing a penalty on model parameters. By far, the most popular model in this group is \emph{Lasso} (least absolute shrinkage and selection operator) proposed by Tibshirani \cite{tibshirani96}. In this approach, the coefficients are obtained by minimizing the residual sum of squares (RSS) subject to a constraint on the magnitude of regression coefficients,
\begin{eqnarray*}
\textrm{minimize } RSS(\beta)  \qquad \\
 \qquad \qquad \textrm{subject to } \sum_{j=1}^{D}|\beta_{j}| \le t 
\end{eqnarray*}
One could estimate the parameters by solving the following optimization problem:
\begin{eqnarray*}
\textrm{minimize } RSS(\beta) + \lambda \sum_{j=1}^{D}|\beta_{j}|
\end{eqnarray*}
where $\lambda \ge 0$ is the regularization parameter. Park and Casella \cite{park08} and Hans \cite{hans09} have proposed a Bayesian alternative method, called Bayesian Lasso. Following the work of \cite{andrews74} and \cite{west87}, the prior distribution used in Bayesian Lasso is expressed as scale mixtures of normal distributions. More specifically, the penalty term is replaced by a prior distribution of the form $P(\beta) \propto \exp(-\lambda |\beta_{j}|) $, which can be represented as a scale mixture of normal distributions \cite{west87}. This leads to a hierarchical Bayesian model with full conditional conjugacy; Therefore, the Gibbs sampler can be used for inference. 

Our proposed method in this paper can directly handle the constraints in Lasso so we can put the commonly used Gaussian prior for model parameters, $\beta|\sigma^2 \sim \mathcal N(0,\sigma^2 I)$, and use Spherical HMC with the transformation discussed in Section \ref{sec:qnorm}. 

We now evaluate our method based on the diabetes data set discussed in \cite{park08}. 
Figure \ref{fig:lasso} compares coefficient estimates given by the Gibbs sampler \cite{park08}, Wall HMC, and Spherical HMC algorithms as the shrinkage factor $s:=\Vert \hat\beta^{\textrm{Lasso}}\Vert_1/\Vert \hat\beta^{\textrm{OLS}}\Vert_1$ changes from 0 to 1. Here, $\hat\beta^{\textrm{OLS}}$ denotes the estimates obtained by ordinary least squares (OLS) regression. For the Gibbs sampler, we choose different $\lambda$ so that the corresponding shrinkage factor $s$ varies from 0 to 1. For Wall HMC and Spherical HMC, we fix the number of leapfrog steps to 10 and set the trajectory length such that they both have comparable acceptance rates around 70\%.

Figure \ref{fig:bridgeff} compares the sampling efficiency of these three methods. As we impose tighter constraints (i.e., lower shrinkage factors), our method becomes substantially more efficient than the Gibbs sampler and Wall HMC. 

\begin{knitrout}
\definecolor{shadecolor}{rgb}{0.969, 0.969, 0.969}\color{fgcolor}\begin{figure}[htbp]

{\centering \includegraphics[width=.7\textwidth,height=.4\textwidth]{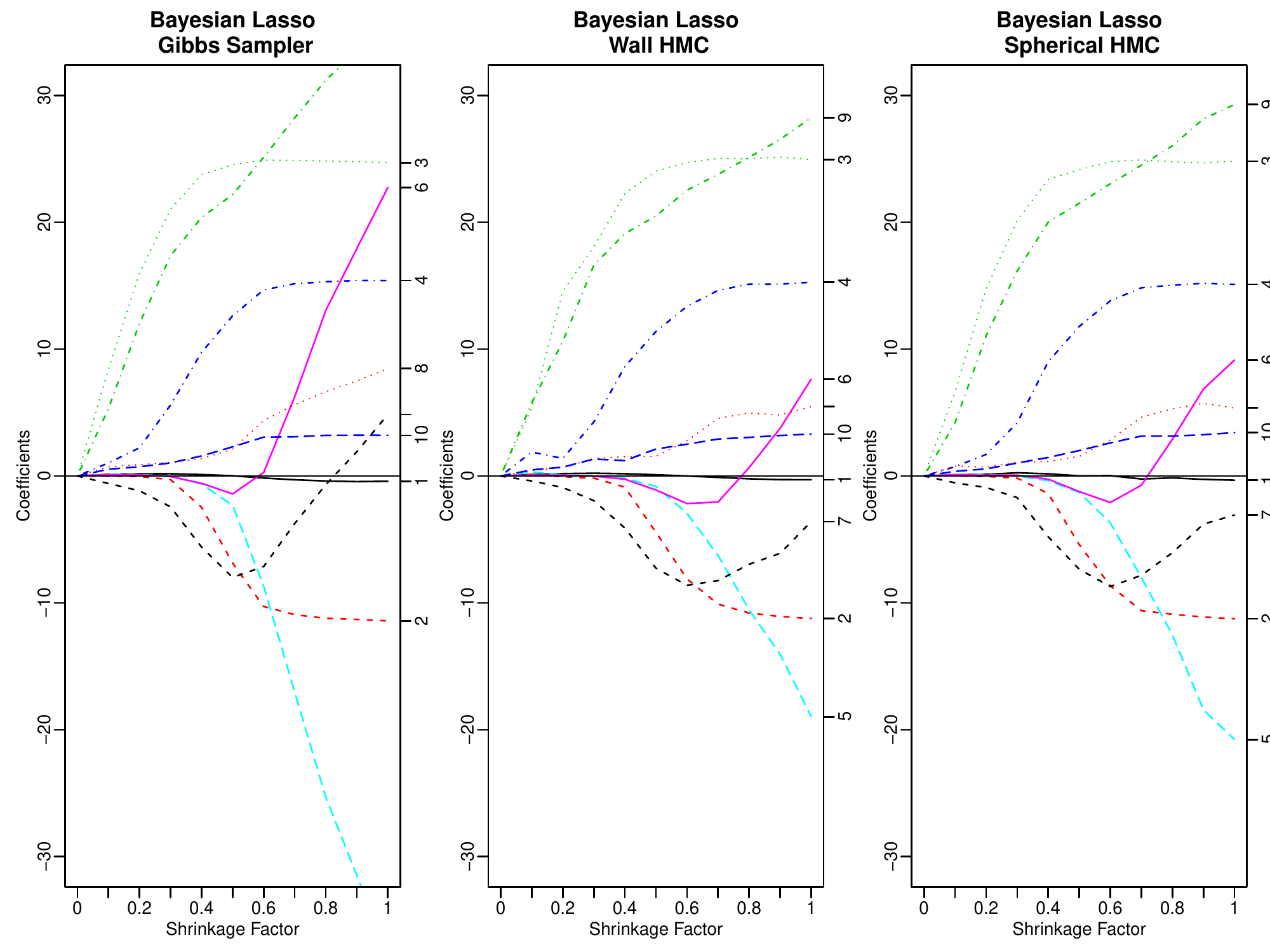} 

}

\caption[Bayesian Lasso using different sampling algorithms]{Bayesian Lasso using three different sampling algorithms: Gibbs sampler (left), Wall HMC (middle) and Spherical HMC (right).\label{fig:lasso}}
\end{figure}

\end{knitrout}

\begin{knitrout}
\definecolor{shadecolor}{rgb}{0.969, 0.969, 0.969}\color{fgcolor}\begin{figure}[htbp]

{\centering \includegraphics[width=.7\textwidth,height=.5\textwidth]{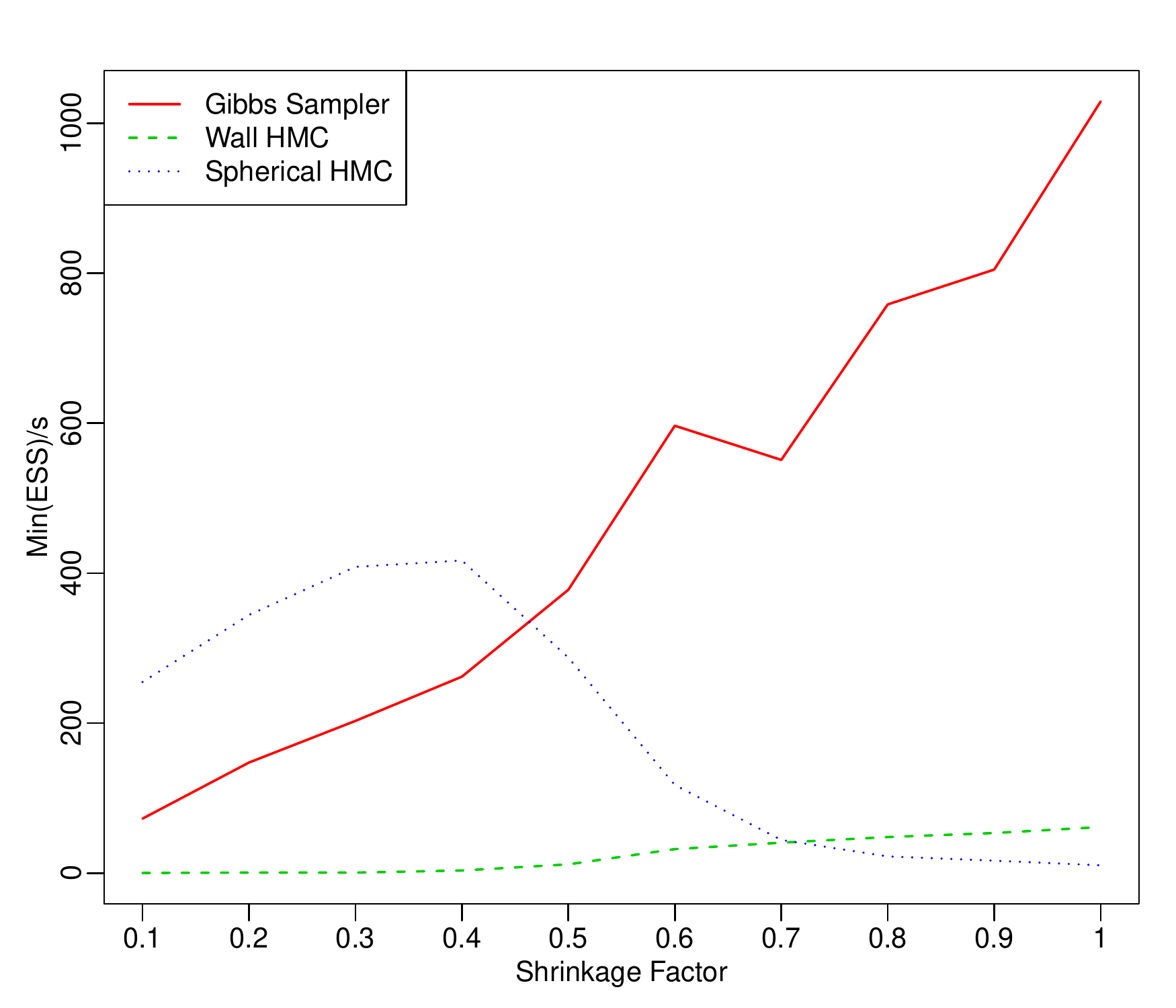} 

}

\caption[Sampling Efficiency in Bayesian Lasso]{Sampling efficiency of different algorithms for Bayesian Lasso based on the diabetes dataset. \label{fig:bridgeff}}
\end{figure}

\end{knitrout}

\subsection{Bridge regression}

The Lasso model discussed in the previous section is in fact a member of a family of regression models called \emph{Bridge regression} \cite{frank93}, where the coefficients are obtained by minimizing the residual sum of squares subject to a constraint on the magnitude of regression coefficients as follows: 
\begin{eqnarray*}
\textrm{minimize } RSS(\beta) =  \sum_{i} (y_{i} - \beta_{0} - x_{i}^{T} \beta)^{2}\\
\textrm{subject to } \sum_{j=1}^{D}|\beta_{j}|^{q} \le t \qquad
\end{eqnarray*}
For Lasso, $q=1$, which allows the model to force some of the coefficients to become exactly zero (i.e., become excluded from the model).

As mentioned earlier, our Spherical HMC method can easily handle this type of constraints through the following transformation:
\begin{equation}
T: \{\beta\in \mathbb R^D:\Vert \beta\Vert_q\leq t\} \rightarrow S^D, \quad \beta_i\mapsto \beta_i' = \beta_i/ t \mapsto \theta_i= \mathrm{sgn}(\beta_i') |\beta_i'|^{q/2},\; \theta \mapsto \tilde\theta = (\theta,\sqrt{1-\Vert \theta\Vert_2^2})
\end{equation}

Figure \ref{fig:bridge} compares the parameter estimates of Bayesian Lasso to the estimates obtained from two Bridge regression models with $q=1.2$ and $q=0.8$ for the diabetes dataset \cite{park08} using our Spherical HMC algorithm. As expected, tighter constraints (e.g., $q=0.8$) would lead to faster shrinkage of regression parameters as we change $s$. 

\begin{knitrout}
\definecolor{shadecolor}{rgb}{0.969, 0.969, 0.969}\color{fgcolor}\begin{figure}[htbp]

{\centering \includegraphics[width=.7\textwidth,height=.4\textwidth]{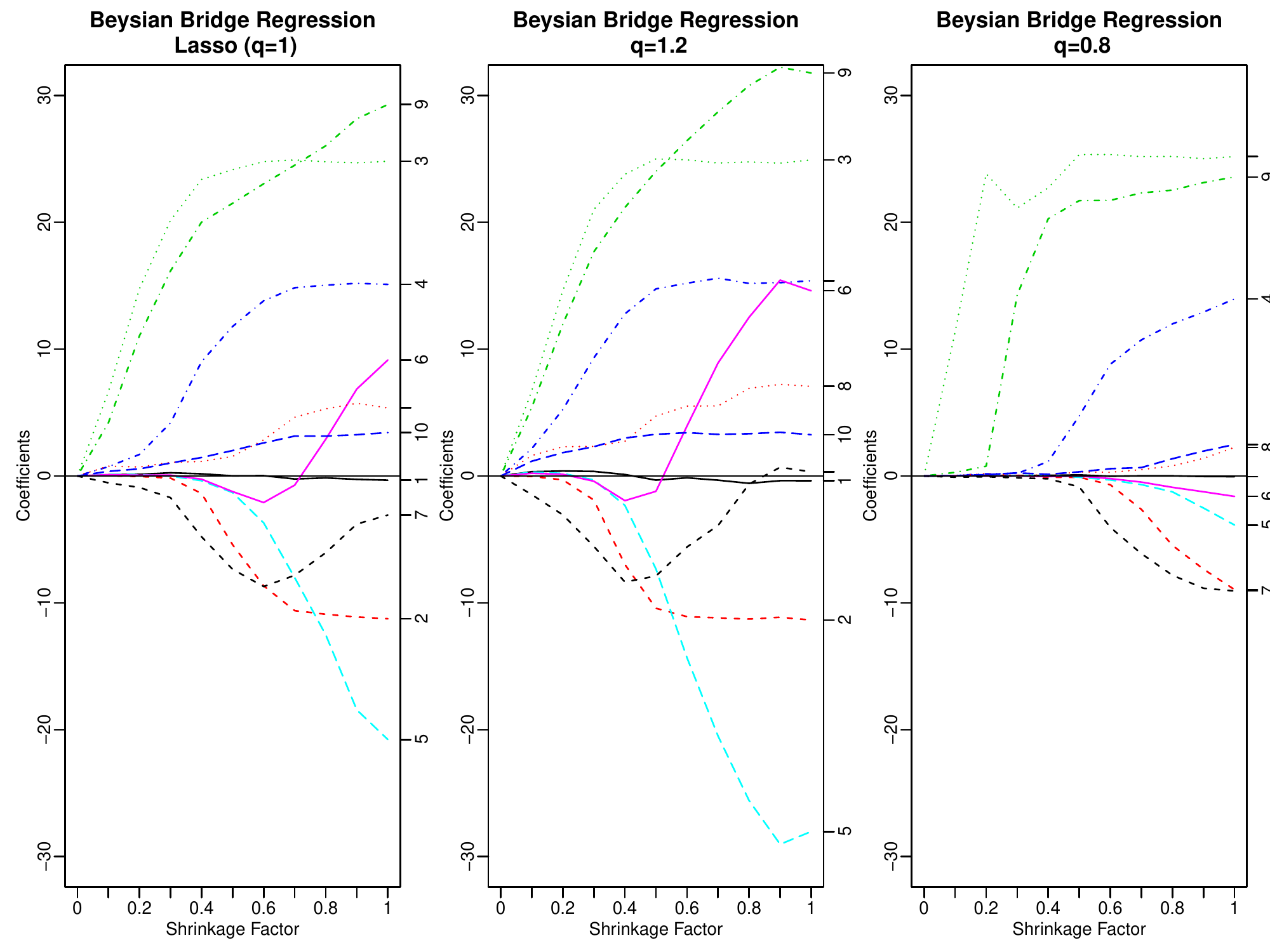} 

}

\caption[Bayesian Bridge Regression by Spherical HMC]{Bayesian Bridge Regression by Spherical HMC: Lasso (q=1, left), q=1.2 (middle), and q=0.8 (right).\label{fig:bridge}}
\end{figure}

\end{knitrout}

\subsection{Modeling synchrony among multiple neurons}

Shahbaba et~al.~\cite{shahbabaSpike} have recently proposed a semiparametric Bayesian model to capture dependencies among multiple neurons by detecting their co-firing patterns over time. In this approach, after discretizing time, there is at most one spike in each interval. The resulting sequence of 1's (spike) and 0's (silence) for each neuron is denoted as $y_{j}$ and is modeled using the logistic function of a continuous latent variable with a Gaussian process prior. For multiple neurons, the corresponding marginal distributions is coupled to their joint probability distribution using a parametric copula model. Let $H$ be $n$-dimensional distribution functions with marginals $F_1,...,F_n$. In genreal, an $n$-dimensional copula is a function of the following form:
\begin{eqnarray*}
H(y_1,...,y_n) = \mathcal{C}(F_1(y_1),...,F_n(y_n)), \ \textrm{for all } y_1, \ldots,y_n
\end{eqnarray*}
Here, $\mathcal{C}$ defines the dependence structure between the marginals. Shahbaba et~al.~\cite{shahbabaSpike} use special case of the Farlie-Gumbel-Morgenstern (FGM) copula family \cite{farlie60, gumbel60, morgenstern56, nelsen98}. For $n$ random variables $Y_{1}, Y_{2}, \ldots, Y_{n}$, the FGM copula, $\mathcal{C}$, has the following form:
\begin{eqnarray*}
\big[1+\sum_{k=2}^n \ \ \sum_{1\le j_1< \cdots <j_k\le n}\beta_{j_1j_2...j_k}\prod_{l=1}^k(1-F_{j_l}) \big] \prod_{i=1}^nF_i 
\end{eqnarray*}
where $F_i=F_i(y_i)$. Restricting the model to second-order interactions, we have
\begin{eqnarray*}\label{copula}
H(y_1, \ldots ,y_n)=\big[1+\sum_{1\le j_1<j_2\le n}\beta_{j_1j_2}\prod_{l=1}^2(1-F_{j_l})\big] \prod_{i=1}^nF_i
\end{eqnarray*}
where $F_i=P(Y_i\le y_i)$. Here, $y_{1}, \ldots, y_{n}$ denote the firing status of $n$ neurons at time $t$; $\beta_{j_1 j_2}$ captures the relationship between the $j_1^{th}$ and $j_2^{th}$ neurons. To ensure that probability distribution functions remain within $[0, 1]$, the following constraints on $n\choose 2$ parameters $\beta_{j_1j_2}$ are imposed:
\begin{equation}\label{constr}
1+\sum_{1\leq j_1<j_2\leq n} \beta_{j_1j_2}\prod_{l=1}^2 \epsilon_{j_l}\geq 0, \quad\epsilon_1,\cdots, \epsilon_n\in\{-1,1\}
\end{equation}
Considering all possible combinbinations of $\epsilon_{j_1}$ and
$\epsilon_{j_2}$ in \eqref{constr}, there are $n(n-1)$ linear inequalities, which can be combined into the following inequality:
\begin{equation}\label{simpineq}
\sum_{1\leq j_1<j_2\leq n} |\beta_{j_1j_2}| \leq 1
\end{equation}
For this model, we can use the square root mapping described in
section \ref{sec:qnorm} to transform such diamond domain of parameters to the unit ball before using Spherical HMC.

\begin{figure}[t]
  \begin{center}
  {\includegraphics[width=50mm,height=30mm]{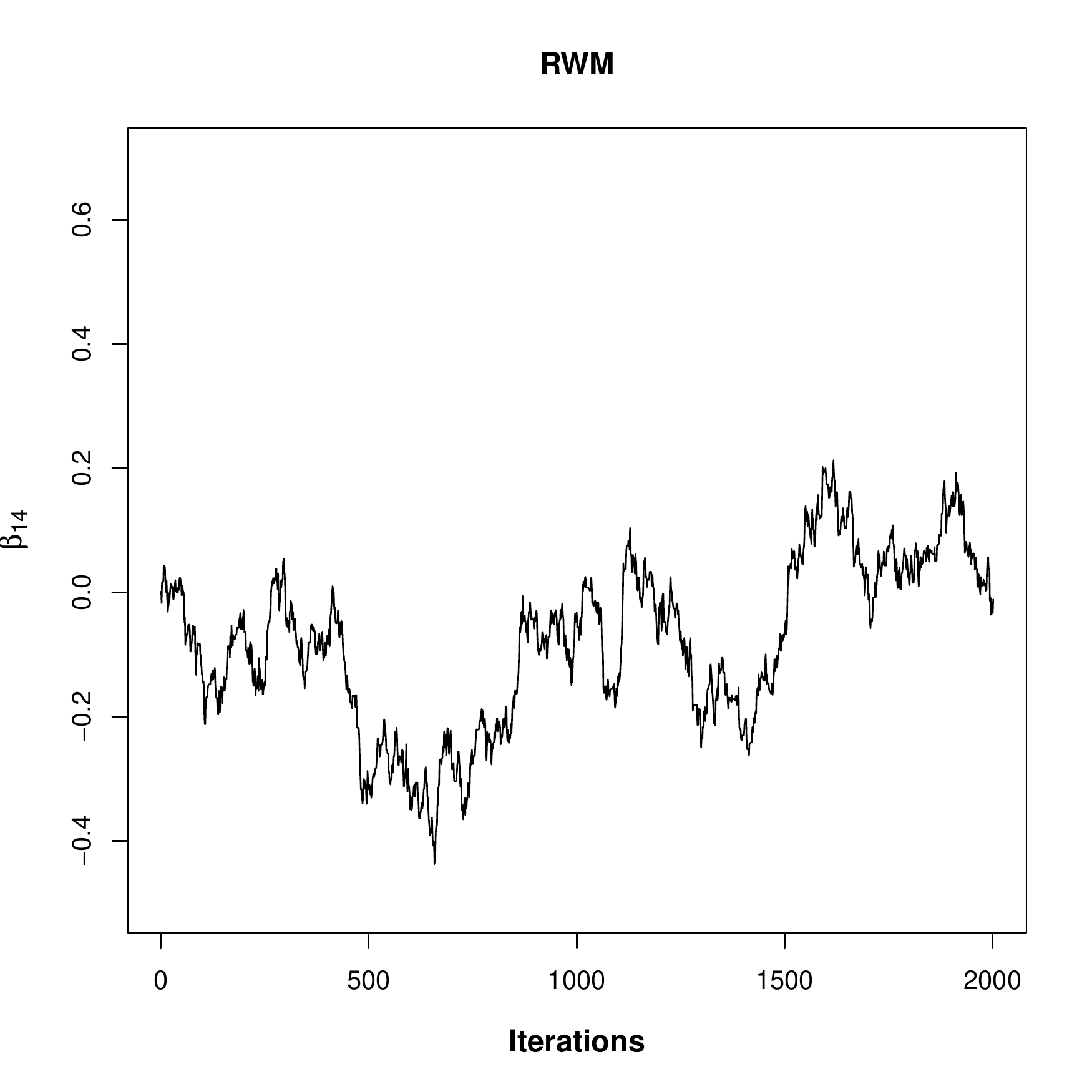}}
  {\includegraphics[width=50mm,height=30mm]{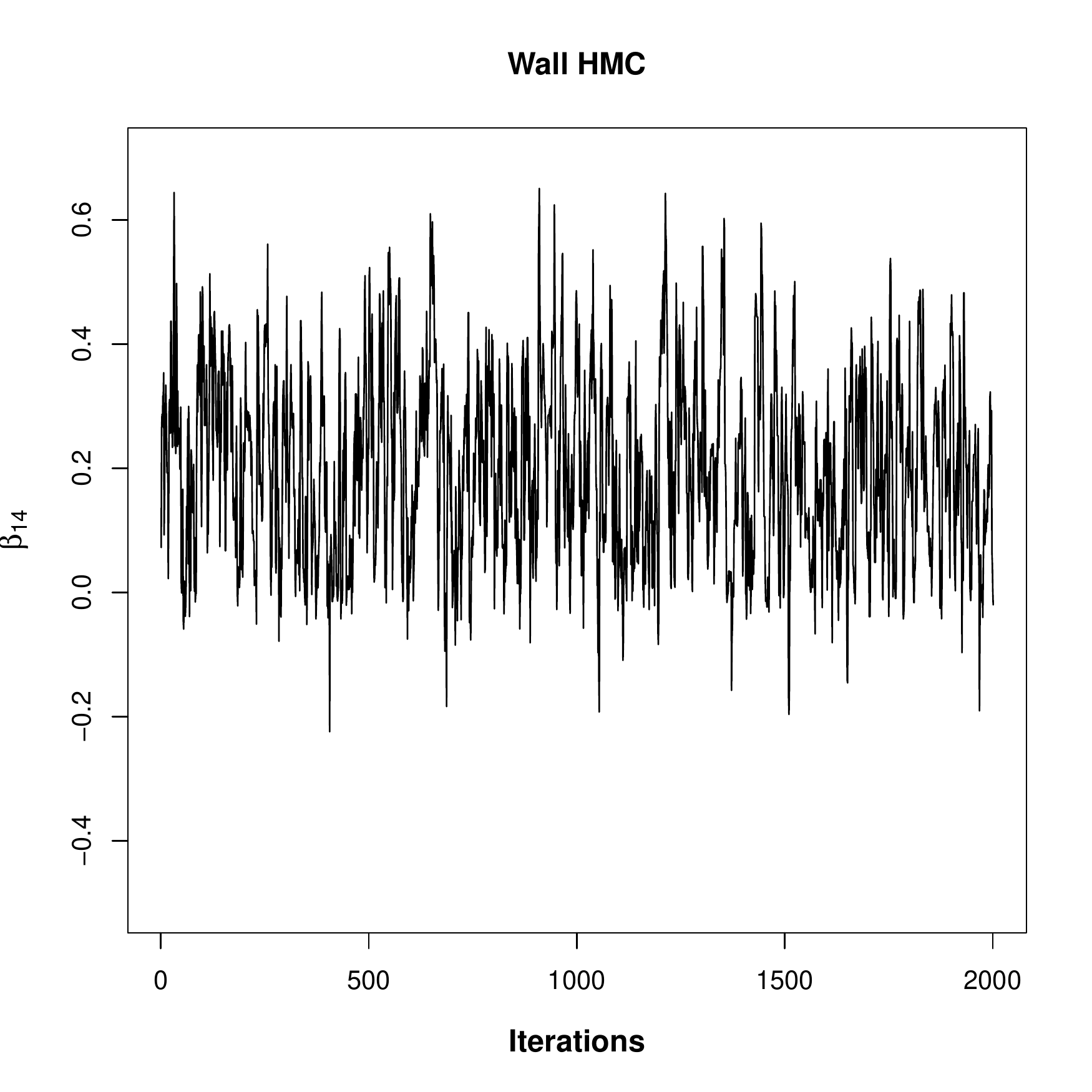}}
  {\includegraphics[width=50mm,height=30mm]{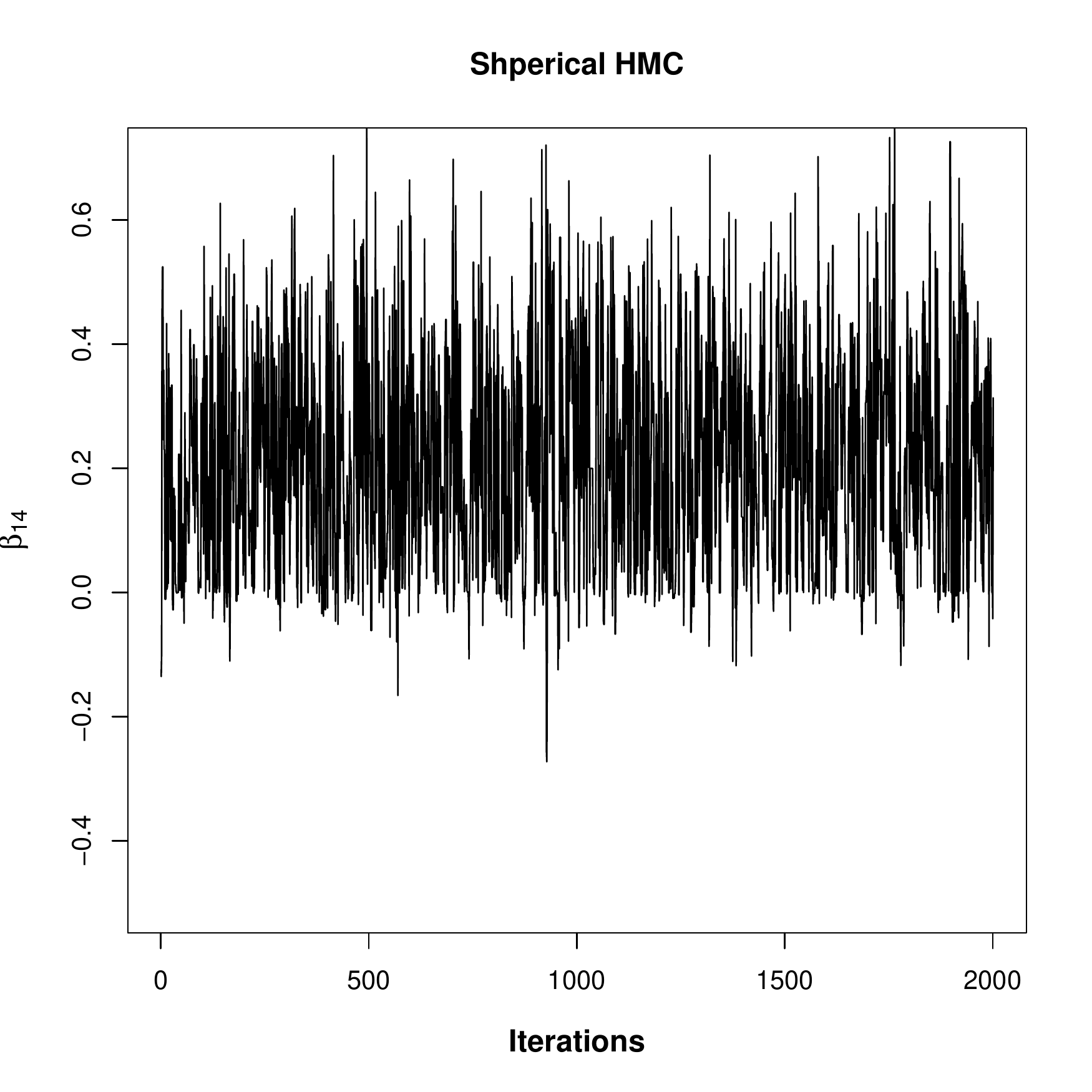}}
  \caption{ Trace Plots of $\beta_{14}$ under the rewarded stimulus. }
  \label{Rewarded}
  \end{center}
  \begin{center}
  {\includegraphics[width=50mm,height=30mm]{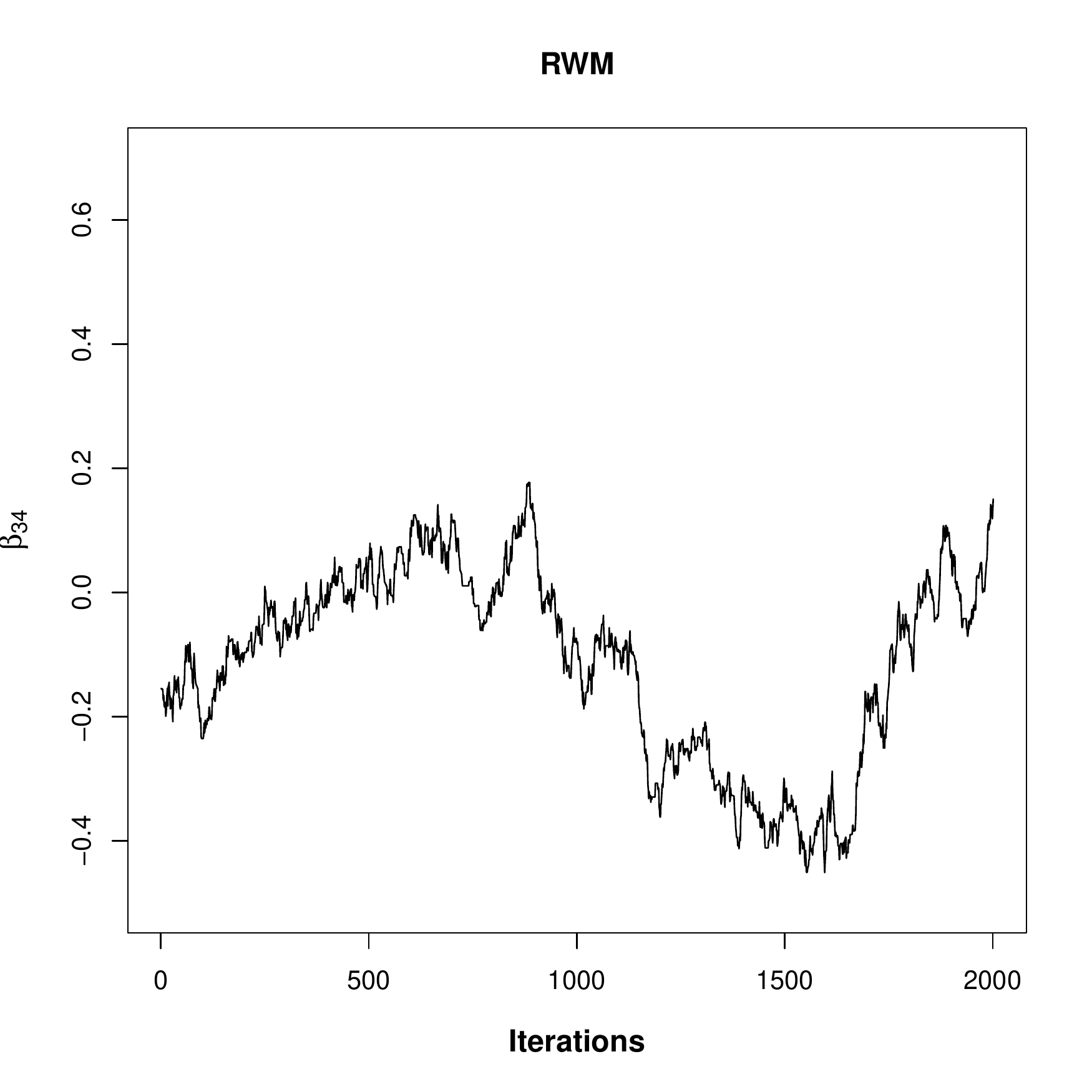}}
  {\includegraphics[width=50mm,height=30mm]{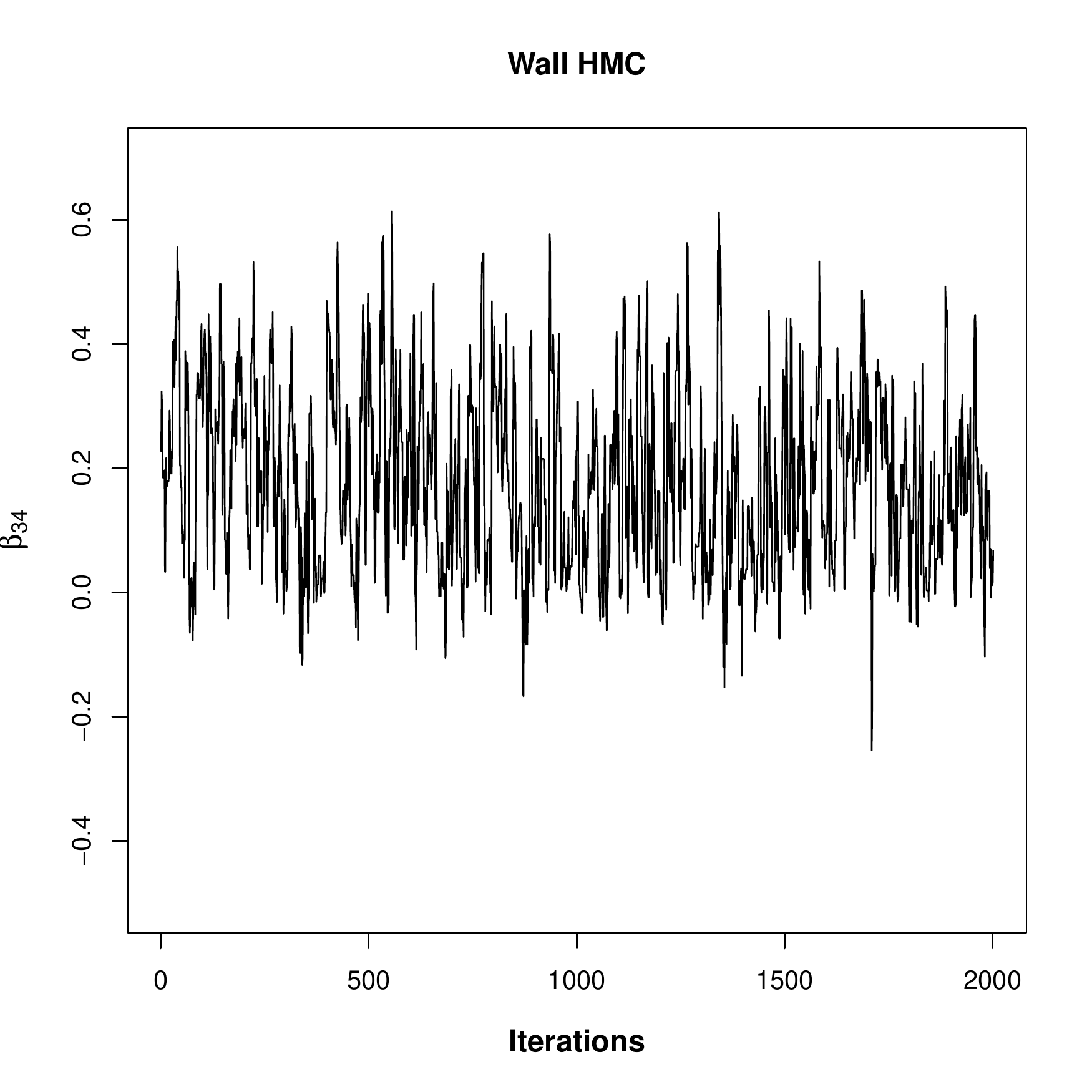}}
  {\includegraphics[width=50mm,height=30mm]{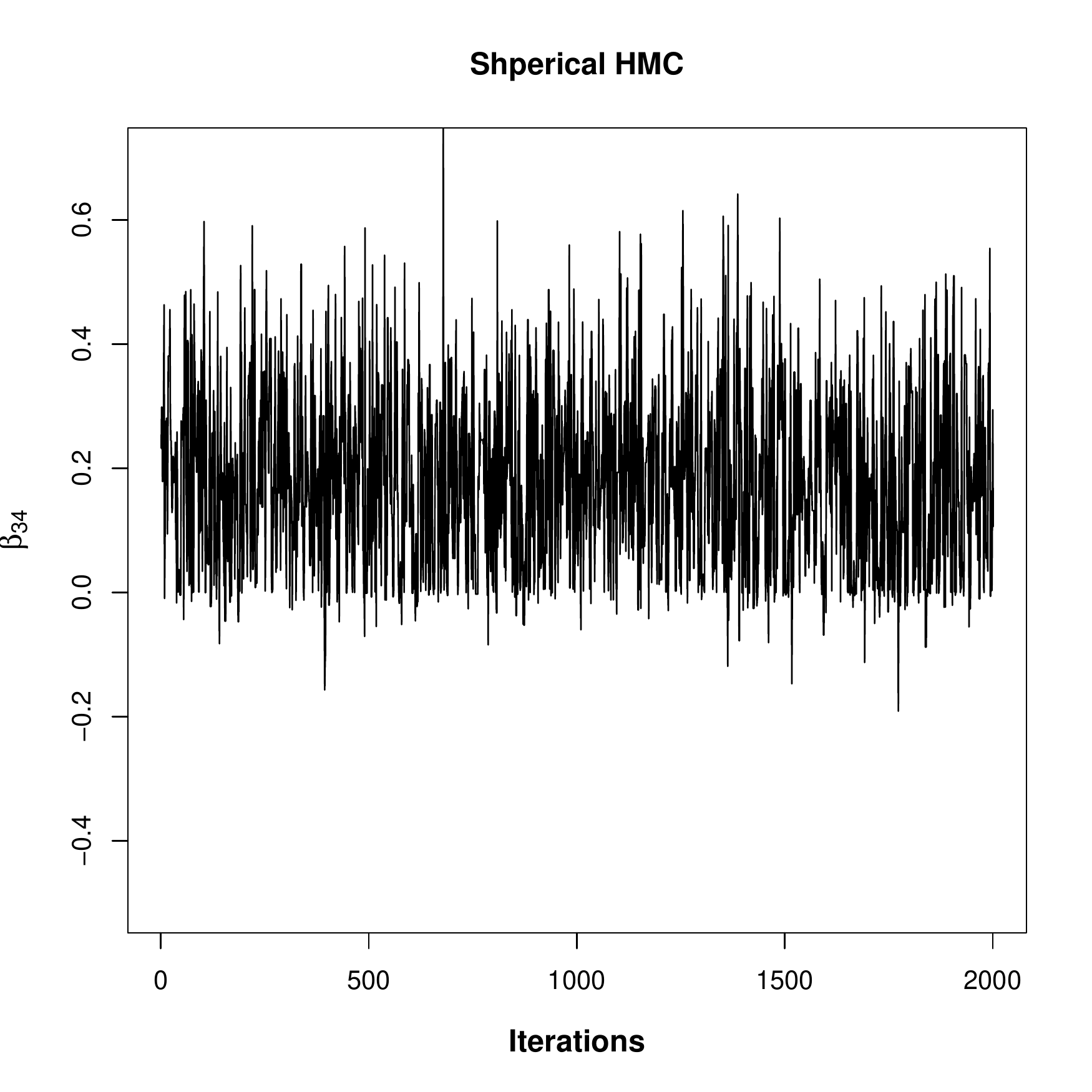}}
  \caption{ Trace Plots of $\beta_{34}$ under the non-rewarded stimulus. }
  \label{nonRewarded}
  \end{center}
\end{figure}

We apply our method to a real dataset based on an experiment investigating the role of prefrontal cortical area in rats with respect to reward-seeking behavior discussed in \cite{shahbabaSpike}. For more details regarding this experiment, see \cite{shahbabaSpike}. Here, we focus on 5 simultaneously recorded neurons. The copula model detected significant associations among three neurons: the $1^{st}$ and $4^{th}$ neurons ($\beta_{14}$) under the rewarded stimulus, and the $3^{th}$ and $4^{th}$ neurons ($\beta_{3,4}$) under the non-rewarded stimulus. All other parameters were deemed non-significant (based on the lower tail probability of zero). The trace plots of $\beta_{14}$ under the rewarded stimulus and $\beta_{34}$ under the non-rewarded stimulus are provided in Figure \ref{Rewarded} and Figure \ref{nonRewarded} respectively. As we can see in these figures and Table \ref{tab:spike-eff}, Spherical HMC is substantially more efficient than RWM and Wall HMC.

\begin{table}[ht]
\centering
\begin{tabular}{l|l|cccc}
  \hline
Scenario & Method & AP & s & ESS & Min(ESS)/s \\ 
  \hline
 & RWM & 0.78 & 2.59 & (6,10,17) & 7.08e-04 \\ 
  Rewarded Stimulus & Wall HMC & 0.91 & 18.25 & (231,319,590) & 4.23e-03 \\ 
   & Spherical
HMC & 0.83 & 17.04 & (1012,1771,2001) & 1.98e-02 \\ 
   \hline
 & RWM & 0.76 & 2.38 & (4,9,21) & 5.74e-04 \\ 
  Non-rewarded Stimulus & Wall HMC & 0.76 & 17.69 & (193,241,409) & 3.63e-03 \\ 
   & Spherical
HMC & 0.81 & 18.03 & (1216,1620,2001) & 2.25e-02 \\ 
   \hline
\end{tabular}
\caption{Comparing sampling efficiencies of RWM, Wall HMC, and Spherical HMC based on the copula model for detecting synchrony among five neurons under rewarded stimulus and non-rewarded stimulus. } 
\label{tab:spike-eff}
\end{table}


\section{Discussion} \label{discussion}

We have introduced a new efficient sampling algorithm for constrained distributions. Our method first maps the parameter space to the unit ball and then augments the resulting space to a sphere. A dynamical system is then defined on the sphere to propose new states that are guaranteed to remain within the boundaries imposed by the constraints. We have also shown how our method can be used for other types of constraints after mapping them to the unit ball. Further, by using the splitting strategy, we could improve the computational efficiency of our algorithm. 

In this paper, we assumed the Euclidean metric ${\bf I}$ on unit ball, ${\bf B}^D_0(1)$. The proposed approach can be extended to more complex metrics, such as the Fisher information metric
${\bf G_F}$, in order to exploit the geometric properties of the parameter space \cite{girolami11}. This way, the metric for the augmented space could be defined as ${\bf G_F} + \theta \theta^T/\theta_{D+1}^2$. Under such a metric however, we might not be able to find the geodesic flow analytically. Therefore, the added benefit from using the Fisher information metric might be undermined by the resulting computational overhead. See \cite{girolami11} and \cite{byrne13} for more discussion.


We have discussed several applications of our method in this current paper. The proposed method of course can be applied to other problems involved constrained target distributions. Further, the ideas presented here can be employed in other MCMC algorithms.


\newpage
\begin{center}
{\huge \bf Appendix: Derivations and Proofs}
\end{center}
\appendix

\section{From unit ball to sphere}\label{geomS}
Consider the $D$-dimensional ball ${\bf B}^D_0(1)=\{\theta\in \mathbb R^D: \Vert
\theta\Vert_2\leq 1\}$ and the $D$-dimensional sphere ${\bf S}^D=\{\tilde\theta=(\theta,\theta_{D+1})\in \mathbb
R^{D+1}: \Vert \tilde\theta\Vert_2= 1\}$. Note that $\{\theta, {\bf B}^D_0(1)\}$ can be viewed as a coordinate chart for ${\bf
S}^D$. For ${\bf S}^D$, the first fundamental formula $ds^{2}$, i.e., squared infinitesimal length of a curve, is explicitly expressed in terms of the differential form $d\theta$ and the \emph{canonical metric} ${\bf G_S}$  as follows:
\begin{eqnarray*}
ds^{2} = \langle d\theta, d\theta \rangle_{\bf G_S} = d\theta^{T}{\bf G_S}d\theta
\end{eqnarray*}
which can be obtained as follows \cite{spivak79-1}:
\begin{equation}
ds^2 = \sum_{i=1}^{D+1} d\theta_i^2 = \sum_{i=1}^D d\theta_i^2 + (d(\theta_{D+1}(\theta)))^2 = d\theta^T d\theta + \frac{(\theta^T d\theta)^2}{1-\Vert \theta\Vert_2^2} = d\theta^T [I + \theta \theta^T/\theta_{D+1}^2] d\theta
\end{equation}
Therefore, the canonical metric ${\bf G_S}$ of ${\bf S}^D$ is
\begin{equation}
{\bf G_S} = I_D + \frac{\theta \theta^T}{\theta_{D+1}^2}
\end{equation}
For any vector $\tilde v=(v,v_{D+1})\in T_{\tilde\theta}{\bf S}^D=\{\tilde v\in \mathbb
R^{D+1}: \tilde\theta^T\tilde v=0\}$, one could view ${\bf G_S}$ as a mean to
express the length of $\tilde v$ in $v$:
\begin{equation}
v^T{\bf G_S}v = \Vert v\Vert_2^2 + \frac{v^T\theta \theta^T v}{\theta_{D+1}^2} = \Vert v\Vert_2^2 + \frac{ (-\theta_{D+1} v_{D+1})^2}{\theta_{D+1}^2} = \Vert v\Vert_2^2 + v_{D+1}^2 = \Vert \tilde v\Vert_2^2
\end{equation}
The determinant of canonical metric ${\bf G_S}$ is given by the matrix
determinant lemma,
\begin{equation}
\det G_{\bf S} = \det (I_D + \frac{\theta \theta^T}{\theta_{D+1}^2}) = 1+ \frac{\theta^T \theta}{\theta_{D+1}^2} = \frac{1}{\theta_{D+1}^2}
\end{equation}
for which the inverse is obtained by Sherman-Morrison-Woodbury formula \cite{golub96}
\begin{equation}
\det G_{\bf S}^{-1} = \left[ I_D + \frac{\theta \theta^T}{\theta_{D+1}^2} \right]^{-1} = I_D - \frac{\theta \theta^T/\theta_{D+1}^2}{1+\theta^T\theta/\theta_{D+1}^2} = I_D - \theta \theta^T
\end{equation}

\subsection{Jacobian Determinant of $T_{{\bf S}\to {\bf B}}$}\label{app:s2b}
Using the volume form \cite{spivak79-1}, we have
\begin{equation}\label{volform}
\int_{{\bf S}_+^D} f(\tilde\theta) d\tilde\theta_{\bf S} = \int_{{\bf B}_0^D(1)} f(\theta) \sqrt{\det G_{\bf S}} d\theta_{\bf B} 
\end{equation}
The transformation $T_{{\bf B}\to {\bf S}}: \theta \mapsto
\tilde\theta=(\theta,\theta_{D+1}=\sqrt{1-\Vert \theta\Vert_2^2})$ bijectively
maps the unit ball ${\bf B}^D_0(1)$ to upper-hemisphere ${\bf S}^D_+$. Using the change of variable theorem, we have
\begin{equation}
\int_{{\bf S}_+^D} f(\tilde\theta) d\tilde\theta_{\bf S} = \int_{{\bf B}_0^D(1)} f(\theta) \left|\frac{d\tilde\theta_{\bf S}}{d\theta_{\bf B}}\right| d\theta_{\bf B} 
\end{equation}
from which we can obtain the Jacobian determinant of $T_{{\bf B}\to {\bf S}}$ as follows:
\begin{equation}\label{detb2s}
\left|\frac{d\tilde\theta_{\bf S}}{d\theta_{\bf B}}\right| = \sqrt{\det G_{\bf S}} = 1/|\theta_{D+1}|
\end{equation}
Therefore, the Jacobian determinant of $T_{{\bf S}\to {\bf B}}$ is $|d T_{{\bf S}\to {\bf B}}| = \left|\frac{d\tilde\theta_{\bf B}}{d\theta_{\bf S}}\right| = |\theta_{D+1}|$.

\subsection{Geodesic}\label{GEODS}
To find the geodesic on a sphere, we need to solve the following equations:
\begin{eqnarray}\label{geodS}
\dot \theta & = & v \label{geodS:1}\\
\dot v & = & -v^T \Gamma v \label{geodS:2}
\end{eqnarray}
for which we need to calculate the Christoffel symbols, $\Gamma$, first. Note that the $(i,j)$-th element of ${\bf G_S}$ is $g_{ij} = \delta_{ij} +
\theta_i\theta_j/\theta_{D+1}^2$, and the $(i,j,k)$-th element of $d{\bf G_S}$
is $g_{ij,k} = (\delta_{ik}\theta_j +
\theta_i\delta_{jk})/\theta_{D+1}^2 + 2\theta_i\theta_j\theta_k/\theta_{D+1}^4$.
Therefore
\[
\begin{split}
\Gamma_{ij}^k & = \frac{1}{2}g^{kl}[g_{lj,i}+g_{il,j}-g_{ij,l}]\\
& = \frac{1}{2}(\delta^{kl}-\theta^k\theta^l)[(\delta_{li}\theta_j +
\theta_l\delta_{ji})/\theta_{D+1}^2 + (\delta_{ij}\theta_l +
\theta_i\delta_{lj})/\theta_{D+1}^2 - (\delta_{il}\theta_j +
\theta_i\delta_{jl})/\theta_{D+1}^2 + 2\theta_i\theta_j\theta_l/\theta_{D+1}^4]\\
& = (\delta^{kl}-\theta^k\theta^l)\theta_l/\theta_{D+1}^2[\delta_{ij}+\theta_i\theta_j/\theta_{D+1}^2]\\
& = \theta_k[\delta_{ij}+\theta_i\theta_j/\theta_{D+1}^2] = [{\bf G_S}\otimes \theta]_{ijk}
\end{split}
\]
Using these results, we can write Equation \eqref{geodS:2} as $\dot v = -v^T {\bf G_S}v\theta = -\Vert\tilde v\Vert_2^2\theta$. Further, we have
\begin{eqnarray}\label{geodSplus}
\dot \theta_{D+1} = & \frac{d}{dt} \sqrt{1-\Vert \theta\Vert_{2}^{2}}  = -\frac{\theta^T}{\theta_{D+1}}\dot \theta & = v_{D+1}\\
\dot v_{D+1} = &  -\frac{d}{dt} \frac{\theta^T v}{\theta_{D+1}}= -\frac{\dot\theta^T v+ \theta^T\dot v}{\theta_{D+1}} + \frac{\theta^T v}{\theta_{D+1}^{2}}\dot\theta_{D+1} & = -\Vert\tilde v\Vert_2^2\theta_{D+1}
\end{eqnarray}
Therefore, we can rewrite the geodesic equations \eqref{geodS:1}\eqref{geodS:2} as
\begin{eqnarray}\label{geodSx}
\dot{\tilde\theta} & = & \tilde v \label{geodSx:1}\\
\dot{\tilde v} & = & -\Vert\tilde v\Vert_2^2 \tilde \theta \label{geodSx:2}
\end{eqnarray}
Multiplying both sides of Equation \eqref{geodSx:2} by $\tilde v^T$ to obtain $\frac{d\Vert \tilde
v\Vert_2^2}{dt}=0$, we can solve the above system of differential equations as follows: 
\begin{eqnarray}
\tilde\theta(t) & = & \tilde\theta(0) \cos(\Vert \tilde v(0)\Vert_{2} t) + \frac{\tilde v(0)}{\Vert \tilde v(0)\Vert_{2}} \sin(\Vert \tilde v(0)\Vert_{2} t)\\
\tilde v(t) & = & -\tilde\theta(0) \Vert \tilde v(0)\Vert_{2} \sin(\Vert \tilde v(0)\Vert_{2} t) + \tilde v(0) \cos(\Vert \tilde v(0)\Vert_{2} t)
\end{eqnarray}

\section{Transformations between different constrained regions}\label{transJ}
Denote the general hyper-rectangle type constrained region as ${\bf R}^D:=\{\beta\in \mathbb R^D: l \leq \beta\leq u \}$.
For transformations $T_{{\bf S}\to {\bf R}}$ and $T_{{\bf S}\to {\bf Q}}$, we can find the Jacobian determinants as follows. First, we note 
\begin{eqnarray}
T_{{\bf S}\to {\bf R}} = T_{{\bf C}\to {\bf R}}\circ T_{{\bf B}\to {\bf
C}}\circ T_{{\bf S}\to {\bf B}}: \tilde\theta\mapsto \theta\mapsto \beta'=\theta \frac{\Vert\theta\Vert_2}{\Vert\theta\Vert_{\infty}}\mapsto \beta=\frac{u-l}{2}\beta'+\frac{u+l}{2}
\end{eqnarray}
The corresponding Jacobian matrices are
\begin{eqnarray}
T_{{\bf B}\to {\bf C}}:\quad \frac{d\beta'}{d\theta^T} & = & \frac{\Vert\theta\Vert_2}{\Vert\theta\Vert_{\infty}}\left[I + \theta \left(\frac{\theta^T}{\Vert\theta\Vert_2^2} - \frac{e_{\arg\max|\theta|}^T}{\theta_{\arg\max|\theta|}}\right)\right]\\
T_{{\bf C}\to {\bf R}}:\quad \frac{d\beta}{d(\beta')^T} & = & \mathrm{diag}(\frac{u-l}{2})
\end{eqnarray}
where $e_{\arg\max|\theta|}$ is a vector with $\arg\max|\theta|$-th element 1 and all others 0.
Therefore, 
\begin{equation}\label{dets2r}
|dT_{{\bf S}\to {\bf R}}| = |dT_{{\bf C}\to {\bf R}}|\, |dT_{{\bf B}\to {\bf
C}}|\, |dT_{{\bf S}\to {\bf B}}| =  \left|\frac{d\beta}{d(\beta')^T}\right| \left|\frac{d\beta'}{d\theta^T}\right| \left|\frac{d\theta_{\bf B}}{d\tilde\theta_{\bf S}}\right| = |\theta_{D+1}| \frac{\Vert\theta\Vert_2^D}{\Vert\theta\Vert_{\infty}^D} \prod_{i=1}^D \frac{u_i-l_i}{2}
\end{equation}
%
Next, we note 
\begin{eqnarray}
T_{{\bf S}\to {\bf Q}} = T_{{\bf B}\to {\bf Q}}\circ T_{{\bf S}\to {\bf B}}: \tilde\theta\mapsto \theta\mapsto \beta=\mathrm{sgn}(\theta)|\theta|^{2/q}
\end{eqnarray}
The Jacobian matrix for $T_{{\bf B}\to {\bf Q}}$ is
\begin{equation}
\quad \frac{d\beta}{d\theta^T} = \frac{2}{q}\mathrm{diag}(|\theta|^{2/q-1})
\end{equation}
Therefore the Jacobian Determinant of $T_{{\bf S}\to {\bf Q}}$ is
\begin{equation}\label{dets2q}
|dT_{{\bf S}\to {\bf Q}}| = |dT_{{\bf B}\to {\bf Q}}|\, |dT_{{\bf S}\to {\bf
B}}| =  \left|\frac{d\beta}{d\theta^T}\right| \left|\frac{d\theta_{\bf B}}{d\tilde\theta_{\bf S}}\right|= \left(\frac{2}{q}\right)^D \left(\prod_{i=1}^{D}|\theta_i|\right)^{2/q-1} |\theta_{D+1}|
\end{equation}

\section{Splitting Hamilton dynamics on ${\bf S}^D$}\label{splitHS}
Although splitting the Hamiltonian function and its usefulness in improving HMC is a well-studied topic of research\cite{leimkuhler04, shahbabaSplitHMC, byrne13}, splitting the Lagrangian function, which is used in our approach, has not been discussed in the literature, to the best of our knowledge. Therefore, we prove the validity of our splitting method by starting with the well-understood method of splitting Hamiltonian \cite{byrne13},
\begin{equation}
H^*(\theta,p) = U(\theta)/2 + \frac{1}{2} p^T {\bf G_S}^{-1} p + U(\theta)/2
\end{equation}
The corresponding systems of differential equations,
\begin{eqnarray}
\left\{\begin{array}{lcl}
\dot \theta & = & 0\\
\dot p & = & -\frac{1}{2} \nabla U(\theta)
\end{array}\right.\label{sphHD:U}
\qquad
\left\{\begin{array}{lcl}
\dot \theta & = & {\bf G_S}^{-1} p\\
\dot p & = & -\frac{1}{2} p^T{\bf G_S}^{-1} d{\bf G_S} {\bf G_S}^{-1}p
\end{array}\right.\label{sphHD:K}
\end{eqnarray}
can be written in terms of Lagrangian dynamics as follows:
in $(\theta, v)$ \cite{lan12}:
\begin{eqnarray}
\left\{\begin{array}{lcl}
\dot \theta & = & 0\\
\dot v & = & -\frac{1}{2} {\bf G}_{\bf S}^{-1} \nabla U(\theta)
\end{array}\right.\label{sphLDa:U}
\qquad
\left\{\begin{array}{lcl}
\dot \theta & = & v\\
\dot v & = & -v^T \Gamma v
\end{array}\right.\label{sphLDa:K}
\end{eqnarray}
We have solved the second dynamics (on the right) in Section \ref{GEODS}. To solve the first dynamics, we note that
\begin{eqnarray}\label{sphLDa:Uplus}
\dot \theta_{D+1} = & -\frac{\theta^T}{\theta_{D+1}}\dot \theta & = 0\\
\dot v_{D+1} = & -\frac{\dot\theta^T v+ \theta^T\dot v}{\theta_{D+1}} + \frac{\theta^T v}{\theta_{D+1}^{2}}\dot\theta_{D+1} & = \frac{1}{2} \frac{\theta^T}{\theta_{D+1}} {\bf G}_{\bf S}^{-1} \nabla U(\theta)
\end{eqnarray}
Therefore, we have
\begin{eqnarray}
\tilde\theta(t) & = & \tilde\theta(0)\\
\tilde v(t) & = & \tilde  v(0) -\frac{t}{2} \begin{bmatrix} I\\ -\frac{\theta(0)^T}{\theta_{D+1}(0)}\end{bmatrix} [I-\theta(0)\theta(0)^T] \nabla U(\theta)
\end{eqnarray}
where $\begin{bmatrix} I\\ -\frac{\theta(0)^T}{\theta_{D+1}(0)}\end{bmatrix} [I-\theta(0)\theta(0)^T]=\begin{bmatrix} I-\theta(0)\theta(0)^T\\ -\theta_{D+1}(0) \theta(0)^T\end{bmatrix} = \begin{bmatrix} I\\ 0\end{bmatrix} - \tilde\theta(0) \theta(0)^T$.
Finally, we note that $\Vert\tilde\theta(t)\Vert_2=1$ if
$\Vert\tilde\theta(0)\Vert_2=1$ and $\tilde v(t)\in T_{\tilde\theta(t)} {\bf
S}^D$ if $\tilde v(0)\in T_{\tilde\theta(0)} {\bf S}^D$.

\end{document}